\def \alphag{{\alpha}^{\mathrm{II}}}
\def \rhot{{\rho}_{\mathrm{t}}}

\def \parx{{\partial_1}}
\def \pary{{\partial_2}}
\def \parphi{{\partial_{\varphi}}}
\def \vel{{\mathrm{v}}}
\def \div{{\mathrm{div}}}
\def \Div{{\mathrm{Div}}}

\documentclass[
    ,final            
  ]
  {aipproc}

\layoutstyle{8x11single}

\begin{document}

\title{Dislocation transport and line length increase in averaged descriptions of dislocations}


\author{T.~Hochrainer}{
  address={%
    Fraunhofer-Institut  f\"ur  Werkstoffmechanik IWM,
    W\"ohlerstr. 11, 79108 Freiburg, Germany
  }
}

\author{M.~Zaiser}{
  address={%
    The University of Edinburgh, Institute for Materials and Processes,
    The Kings Buildings, Sanderson Building, Edinburgh EH9 3JL, United Kingdom
  }
}

\author{P.~Gumbsch}{
  address={%
    Institut f\"ur Zuverl\"assigkeit von Bauteilen und Systemen,
    Universit\"at Karlsruhe (TH), Kaiserstr. 12, 76131 Karlsruhe, Germany
  },
  email={peter.gumbsch@kit.edu}
}

\begin{abstract}
Crystal plasticity is the result of the motion and interaction of dislocations. There is, however, still a major gap between microscopic and mesoscopic simulations and continuum crystal plasticity models. Only recently a higher dimensional dislocation density tensor was defined which overcomes some drawbacks of earlier dislocation density measures. The evolution equation for this tensor can be considered as a continuum version of dislocation dynamics. We use this evolution equation to develop evolution equations for the total dislocation density and an average curvature which together govern a faithful representation of the dislocation kinematics without having to use extra dimensions.\\

\end{abstract}

\maketitle


\section{Introduction}

Dislocation based continuum modelling of plasticity started in the 1950s with the works of Nye \cite{nye53}, Kr\"oner \cite{kroener58}, Kondo \cite{kondo52} and Bilby et al.\ \cite{bilby_bs55}, who introduced largely equivalent tensorial descriptions of the dislocation state of a crystal. It was soon recognised that these concepts were not suited as foundations of a plasticity theory. This, together with their reasonably high mathematical prerequisites, condemned these approaches to a rather shadowy existence since ever. They were revived in the 1990s after Fleck, Ashby et al.\ \cite{fleck_etal94} recognised the role of the Kr\"oner tensor in (size-dependent) strain gradient theories as a measure of so-called geometrically necessary dislocations \cite{ashby70} (GND). Since then various strain gradient dependent plasticity theories have been developed, which are sometimes combined with evolution equations for the so-called statistically stored dislocations (e.g.\ \cite{varadhan_bf06} \cite{roters_r06}) (SSD). The latter evolution equations are usually based on a second direction of dislocation based modelling which traces back to Kocks \cite{kocks76}. He was the first to introduce phenomenological evolution laws for the (scalar) total dislocation density. However, the available evolution equations for SSDs are usually driven by plastic strain and local in the sense that they do not account for dislocation fluxes. This may be justified for polycrystals where dislocations mostly accumulate strain driven because they are trapped at grain boundaries. On the single crystal level, however, plastic strain accumulates where dislocations have passed through, while the dislocation density accumulates where dislocations get trapped, i.e.\ where plastic strain is low.

In \cite{hochrainer_zg07} we showed that dislocation fluxes and the evolution of plastic strain can be handled consistently by using a higher dimensional dislocation density measure. In the current work we show how the extra dimension may be eliminated from the evolution equations by integration to yield evolution equations for the total dislocation density and the average curvature. These evolution equations are able to handle dislocation fluxes and line length changes without using extra dimensions.

\section{Extended continuum theory of dislocations}

At the heart of the extended continuum theory of dislocations developed by the authors \cite{hochrainer_zg07} lies the so called dislocation density tensor of second order (SODT) $ \alphag $. This tensor is a natural generalisation of the classical dislocation density tensor to a higher dimensional configuration space. It is closely related to the phase space densities of dislocations as introduced by El-Azab \cite{elazab00}.

If dislocations move by glide only, the SODT is defined on the configuration space $ Q = M \times \bigcup_\beta S^1_\beta $, where $ M $ denotes the (spatial) crystal manifold, $ \beta $ indicates the slip systems and the $ S^1_\beta $ are unit circles of directions in the respective glide planes. A point in $ Q $ is considered as composed of a spatial point $ p $ and the angle $ \varphi_\beta $ between a direction in the glide plane and the Burgers vector $ b_\beta $. We may consider the SODT as a sum of tensors defined for each slip system, i.e.\ $ \alpha^\mathrm{II} = \sum_\beta \alpha^\mathrm{II}_\beta $. On each slip system, the SODT is defined by a density function $ \rho_\beta(p,\varphi_\beta) $ giving the average number (per unit area) of dislocations at $ p $ with line direction $ l_\beta(\varphi_\beta) $, and a curvature function $ k_\beta(p,\varphi_\beta) $ characterising the average curvature of these dislocations. In the following we restrict ourselves to a single slip system (and skip the index $\beta$) and chose a coordinate system such that the slip plane is the 1-2 plane and the Burgers vector points in 1-direction: $ b=b^1\parx $. The canonical line direction consequently reads $ l(\varphi) =(\cos \varphi, \sin \varphi) $. The second order dislocation density tensor then takes the form
\begin{equation}
 	\alpha^\mathrm{II} = \rho (p,\varphi) (\cos \varphi, \sin \varphi, k(p,\varphi)) \otimes b.
\end{equation}
We call $ L = (\cos \varphi, \sin \varphi, k(p,\varphi)) $ the generalised line direction.
Because dislocations do not end inside a crystal the SODT needs to be solenoidal, that is
\begin{equation}
 	\Div \; \alpha^\mathrm{II} = (\cos \varphi \parx \rho (p,\varphi) + \sin \varphi \parx \rho (p,\varphi) + \parphi (\rho k(p,\varphi))) \otimes b =0,
\end{equation}
where $ \Div $ denotes the divergence operator on the configuration space. We note that the last equation is equivalent to $ \Div (\rho L ) = 0$.  Note that in the sequel we will usually drop the arguments of the involved objects to maintain readability.

We assume that the dislocations move in the direction of the spatial velocity field $ \vel = (\vel^1,\vel^2):=v (p,\varphi) (\sin \varphi, -\cos \varphi) $ perpendicular to the line direction $ l $. Furthermore we introduce the generalised velocity $ V = ( \vel^1, \vel^2, \nabla_L v ) $, where the third component $ \nabla_L v $ reflects the rotational velocity of a moving dislocation segment. The evolution equation of the SODT may be given as evolution equations for $ \rho $ and $ k $ (cf. \cite{hochrainer_zg07}) which read
\begin{eqnarray} \label{evolution_rhok}
	\partial_t \rho &=& -\Div (\rho V ) + \rho v  = - \div (\rho \vel ) + \parphi (\rho \nabla_L v) + \rho v k \label{evolution_rho} \quad \textrm{and} \\
	\partial_t k &=& -vk^2 - \nabla_L \nabla_L v + \nabla_V  k.\label{evolution_k}
\end{eqnarray}
For later use we note that instead of the evolution equation for $ k $ we may also look at the evolution of the product $ \rho k $. The evolution equation for $ \rho k $ is easily obtained from (\ref{evolution_rho}) and (\ref{evolution_k}) as
\begin{eqnarray}
 	\partial_t (\rho k) &=& - \Div \left( \rho k V \right) - \Div \left(\nabla_L v \cdot \rho L \right),
\end{eqnarray}
under consideration of the solenoidality of $ \alphag $, i.e.\ of $ \Div \left(\rho L \right) = 0 $.

\subsection{Relation to the classical dislocation density measures}

The total dislocation density $ \rhot $ can be determined from $ \alphag $ as
\begin{equation}
    \rhot =  \int_{0}^{2\pi} \rho \left( \varphi \right) d\varphi.
\end{equation}
The classical disclocation density tensor $ \alpha $ is obtained by integrating the density against the line direction, i.e.\ through
\begin{equation}
    \alpha =  \int_{0}^{2\pi} \rho \left( \varphi \right) (\cos \varphi, \sin \varphi) d\varphi \otimes b.
\end{equation}
As we consider only one Burgers vector we can decompose the dislocation density tensor into $ \alpha = \kappa \otimes b $ with a vector
\begin{equation}
    \kappa =  (\kappa^1, \kappa^2) := \int_{0}^{2\pi} \rho \left( \varphi \right) (\cos \varphi, \sin \varphi) d\varphi.
\end{equation}


\section{Continuum theory of dislocations without extra dimensions}

Under the assumption that the (scalar) velocity $ v $ does not depend on the dislocation line direction, the evolution of the total dislocation density is easily obtained from Eqn. (\ref{evolution_rho}) as
\begin{eqnarray} \label{Eq: Evolution rhot}
    \partial_t \rhot &=&  \int_{0}^{2\pi} \left( -\div (\rho \vel ) - \parphi (\rho \nabla_L (v)) + \rho v k \right) d\varphi \\
	&=&  -\div \left( v \int_{0}^{2\pi} \rho ( \sin \varphi, -\cos \varphi ) d\varphi \right)+  v \int_{0}^{2\pi} \rho k d\varphi \\
        &=&  -\div \left( v (\kappa^2, -\kappa^1) \right) + v \int_{0}^{2\pi} \rho k d\varphi. \label{Eq: Evolution rhot 3}
\end{eqnarray}
The first term on the right hand side measures the net dislocation flux. That this is the right flux term to describe the change in total dislocation density can be seen by realising that dislocations with opposite line directions also move in opposite directions. Therefore only the difference between the amount of dislocations of each `sign' leads to a change of total line length in a volume. The second term accounts for the change of total line length connected to the expansion or shrinkage of curved lines.

As in the extended continuum theory the evolution equation for the total dislocation density needs to be accompanied by an evolution equation for the (average) curvature $ \bar{k} $ in order to capture the line length changes correctly. We make the following definitions:
\begin{eqnarray}
 	\bar{ \rho k } &:=& \int_{0}^{2\pi} \rho k d\varphi \\
	\bar{k} &:=& \frac{\bar{ \rho k }}{\rhot} = \frac{\int_{0}^{2\pi} \rho k d\varphi}{\int_{0}^{2\pi} \rho d\varphi}
\end{eqnarray}
With these definitions and the additional notation $ \kappa^\perp = (\kappa^2, -\kappa^1) $ we rewrite Eqn. (\ref{Eq: Evolution rhot 3}) as
\begin{equation} \label{Eq: Evolution rhot final}
    \partial_t \rhot =  -\div \left( v \kappa^\perp \right) + v \bar{ \rho k } = -\div \left( v \kappa^\perp \right) + v \rhot \bar{k}.
\end{equation}

To derive a closed set of evolution equations we first look at the evolution equation for the averaged product $ \bar{\rho k} $. Note that as in Eqn. (\ref{Eq: Evolution rhot}) the directional part of the generalised divergence $ \Div $ vanishes upon integration over $ \varphi $ and we find
\begin{eqnarray} \label{Eq: evolution rhok}
 	\partial_t \bar{ \rho k } &=& - \int_{0}^{2\pi} \left( \div (\rho k \vel) + \div (\nabla_L v \cdot \rho l ) \right) d\varphi \\
	&=& -  \div \left( v \int_{0}^{2\pi} \rho k ( \sin \varphi, -\cos \varphi ) d\varphi \right) - \div \left( \int_{0}^{2\pi} \nabla_L v \cdot \rho (\cos \varphi, \sin \varphi) d\varphi \right).
\end{eqnarray}
In the evolution equation for $ \rhot $ the divergence term could be expressed in known quantities under the assumption that the velocity $ v $ does not depend on the line direction. In Eqn. (\ref{Eq: evolution rhok}) we can close the first divergence term by assuming that also the curvature is the same for dislocations of each direction. This is a reasonable assumption in quasi-static situation where the curvature essentially balances the local shear stress. With this assumption we find for the first divergence term 
\begin{equation} 
	\div \left( v \int_{0}^{2\pi} \rho k ( \sin \varphi, -\cos \varphi ) d\varphi \right) = \div\left( v \bar{k} \kappa^\perp \right).
\end{equation}
for the second divergence term in Eqn. (\ref{Eq: evolution rhok}) we need a further simplifying assumption. Here we assume that the dislocation density is nearly independent of the orientation which implies that the amount of GND is small compared to the total dislocation density, $ || \kappa || << \rhot $. Note, however, that a small amount of GNDs does not imply an equal distribution of density in the orientation space. We think that a nearly isotropic orientation distribution is a common situation when dealing with mesoscopic average volumes. In this case $ \rho $ is largely independent of $ \varphi $ and we necessarily have $ \rho (\varphi) \cong \rhot / 2 \pi $. In the following we assume that the latter holds as equality. Furthermore we again assume $ v $ to be isotropic and find for the second divergence term
\begin{eqnarray} 
	\div \left( \int_{0}^{2\pi} \nabla_L v \cdot \rho (\cos \varphi, \sin \varphi) d\varphi \right) &=& \div \left(  \frac{\rhot}{2 \pi} \int_{0}^{2\pi} (\cos \varphi \parx v + \sin \varphi \pary v) (\cos \varphi, \sin \varphi) d\varphi \right) \\
	&=& \div \left( \frac{1}{2} \rhot (\parx v, \pary v) \right), 
\end{eqnarray}
where we used
\begin{equation}
 	\int_0^{2\pi} \cos^2(\varphi) d\varphi = \int_0^{2\pi} \sin^2(\varphi) d\varphi = \pi \quad \textrm{and} \quad \int_0^{2\pi} \cos (\varphi)\sin \varphi d\varphi =0.
\end{equation}

In the case of $ v, \rho $ and $ k $ each being (nearly) isotropic we consequently have
\begin{eqnarray} \label{Eq: evolution rhok final}
 	\partial_t \bar{ \rho k } = - \div \left( v \bar{k} \kappa^\perp \right) - \frac{1}{2} \div \left( \rhot \nabla v \right).
\end{eqnarray}
As in the case of the extended continuum theory we may alternatively look at the evolution of $ \bar{k} $ which we can now easily determine from its definition and basic differential calculus as
\begin{eqnarray} \label{Eq: evolution bar k}
 	\partial_t \bar{ k } &=& \partial_t \left( \frac{\bar{ \rho k }}{\rhot} \right) \\
                             &=& - v \bar{k}^2  - \frac{1}{\rho} \left( v \nabla_{\kappa^\perp} \bar{k} + \frac{1}{2} \nabla_{\nabla v} \rhot \right) - \frac{1}{2} \Delta v. \label{Eq: evolution bar k 2}
\end{eqnarray}
Here $ \Delta $ denotes the two-dimensional Laplace operator.

The evolution equation for the dislocation density tensor was derived in \cite{hochrainer_zg07} and translates to the evolution of $ \kappa $ as
\begin{eqnarray} \label{Eq: evolution kappa}
 	\partial_t \kappa &=& (\pary (\rhot v ), - \parx (\rhot v)).
\end{eqnarray}

Under the premise that the assumption of isotropic $ \rho $, $ v $ and $k $ holds, Eqns. (\ref{Eq: Evolution rhot final}), (\ref{Eq: evolution bar k 2}), and (\ref{Eq: evolution kappa}) thus define a kinematically closed system of equations for the evolution of an averaged dislocation system. Note that a fully closed system would need a relation between $ v $, the applied stress and the current dislocation state.

\section{Summary and outlook}

In the current work we used a higher dimensional continuum theory of dislocations to derive coupled evolution equations for the total dislocation density, the classical dislocation density tensor and the average curvature of dislocations. The most important novelty in these equations is that they take into account the change of total dislocation density due to dislocation fluxes. To describe the evolution of the average curvature several simplifying assumptions are needed which we seek to weaken in future work.





\bibliographystyle{aipproc}   

\bibliography{../../references_thomas}

\IfFileExists{\jobname.bbl}{}
 {\typeout{}
  \typeout{******************************************}
  \typeout{** Please run "bibtex \jobname" to optain}
  \typeout{** the bibliography and then re-run LaTeX}
  \typeout{** twice to fix the references!}
  \typeout{******************************************}
  \typeout{}
 }

\end{document}